\documentclass[twocolumn]{aastex631}
\usepackage{multirow}

\newcommand{\fermi}{4FGL J2108.0+5155}
\newcommand{\lhaaso}{LHAASO J2108+5157}
\newcommand\fermilat{\emph{Fermi}-LAT\ }
\newcommand{\xmm}{\textit{XMM-Newton}}

\usepackage{gensymb}
\usepackage{soul}
\usepackage{amsmath}
\usepackage{amssymb}
\usepackage{float}

\begin{document}

\title{HAWC, VERITAS, Fermi-LAT and XMM-Newton follow-up observations of the unidentified ultra-high-energy gamma-ray source LHAASO J2108+5157}

\correspondingauthor{Sajan Kumar}
\email{skumar86@umd.edu}

\correspondingauthor{Xiaojie Wang}
\email{xwang32@mtu.edu}

\correspondingauthor{Michael Martin}
\email{mi.martin@utah.edu}

\correspondingauthor{Jooyun Woo}
\email{jw3855@columbia.edu}

\author[0000-0002-9021-6192]{C.~B.~Adams}\email{}\affiliation{Physics Department, Columbia University, New York, NY 10027, USA}
\author[0000-0002-3886-3739]{P.~Bangale}\email{}\affiliation{Department of Physics, Temple University, Philadelphia, PA 19122, USA}
\author[0000-0003-2098-170X]{W.~Benbow}\email{}\affiliation{Center for Astrophysics $|$ Harvard \& Smithsonian, Cambridge, MA 02138, USA}
\author[0000-0001-6391-9661]{J.~H.~Buckley}\email{}\affiliation{Department of Physics, Washington University, St. Louis, MO 63130, USA}
\author[0009-0001-5719-936X]{Y.~Chen}\email{}\affiliation{Department of Physics and Astronomy, University of California, Los Angeles, CA 90095, USA}
\author{J.~L.~Christiansen}\email{}\affiliation{Physics Department, California Polytechnic State University, San Luis Obispo, CA 94307, USA}
\author{A.~J.~Chromey}\email{}\affiliation{Center for Astrophysics $|$ Harvard \& Smithsonian, Cambridge, MA 02138, USA}
\author{M.~Escobar~Godoy}\email{}\affiliation{Santa Cruz Institute for Particle Physics and Department of Physics, University of California, Santa Cruz, CA 95064, USA}
\author{S.~Feldman}\email{}\affiliation{Department of Physics and Astronomy, University of California, Los Angeles, CA 90095, USA}
\author[0000-0001-6674-4238]{Q.~Feng}\email{}\affiliation{Department of Physics and Astronomy, University of Utah, Salt Lake City, UT 84112, USA}
\author[0000-0002-2944-6060]{J.~Foote}\email{}\affiliation{Department of Physics and Astronomy and the Bartol Research Institute, University of Delaware, Newark, DE 19716, USA}
\author[0000-0002-1067-8558]{L.~Fortson}\email{}\affiliation{School of Physics and Astronomy, University of Minnesota, Minneapolis, MN 55455, USA}
\author[0000-0003-1614-1273]{A.~Furniss}\email{}\affiliation{Santa Cruz Institute for Particle Physics and Department of Physics, University of California, Santa Cruz, CA 95064, USA}
\author[0000-0002-0109-4737]{W.~Hanlon}\email{}\affiliation{Center for Astrophysics $|$ Harvard \& Smithsonian, Cambridge, MA 02138, USA}
\author[0000-0003-3878-1677]{O.~Hervet}\email{}\affiliation{Santa Cruz Institute for Particle Physics and Department of Physics, University of California, Santa Cruz, CA 95064, USA}
\author[0000-0001-6951-2299]{C.~E.~Hinrichs}\email{}\affiliation{Center for Astrophysics $|$ Harvard \& Smithsonian, Cambridge, MA 02138, USA and Department of Physics and Astronomy, Dartmouth College, 6127 Wilder Laboratory, Hanover, NH 03755 USA}
\author[0000-0002-6833-0474]{J.~Holder}\email{}\affiliation{Department of Physics and Astronomy and the Bartol Research Institute, University of Delaware, Newark, DE 19716, USA}
\author{Z.~Hughes}\email{}\affiliation{Department of Physics, Washington University, St. Louis, MO 63130, USA}
\author[0000-0002-1432-7771]{T.~B.~Humensky}\email{}\affiliation{Department of Physics, University of Maryland, College Park, MD, USA and NASA GSFC, Greenbelt, MD 20771, USA}
\author[0000-0002-1089-1754]{W.~Jin}\email{}\affiliation{Department of Physics and Astronomy, University of California, Los Angeles, CA 90095, USA}
\author[0000-0002-3638-0637]{P.~Kaaret}\email{}\affiliation{Department of Physics and Astronomy, University of Iowa, Van Allen Hall, Iowa City, IA 52242, USA}
\author{M.~Kertzman}\email{}\affiliation{Department of Physics and Astronomy, DePauw University, Greencastle, IN 46135-0037, USA}
\author{M.~Kherlakian}\email{}\affiliation{Fakult\"at f\"ur Physik \& Astronomie, Ruhr-Universit\"at Bochum, D-44780 Bochum, Germany}
\author[0000-0003-4785-0101]{D.~Kieda}\email{}\affiliation{Department of Physics and Astronomy, University of Utah, Salt Lake City, UT 84112, USA}
\author[0000-0002-4260-9186]{T.~K.~Kleiner}\email{}\affiliation{DESY, Platanenallee 6, 15738 Zeuthen, Germany}
\author[0000-0002-4289-7106]{N.~Korzoun}\email{}\affiliation{Department of Physics and Astronomy and the Bartol Research Institute, University of Delaware, Newark, DE 19716, USA}
\author[0000-0002-5167-1221]{S.~Kumar}\email{}\affiliation{Department of Physics, University of Maryland, College Park, MD, USA }
\author[0000-0003-4641-4201]{M.~J.~Lang}\email{}\affiliation{School of Natural Sciences, University of Galway, University Road, Galway, H91 TK33, Ireland}
\author[0000-0003-3802-1619]{M.~Lundy}\email{}\affiliation{Physics Department, McGill University, Montreal, QC H3A 2T8, Canada}
\author[0000-0001-9868-4700]{G.~Maier}\email{}\affiliation{DESY, Platanenallee 6, 15738 Zeuthen, Germany}
\author[0000-0001-7106-8502]{M.~J.~Millard}\email{}\affiliation{Department of Physics and Astronomy, University of Iowa, Van Allen Hall, Iowa City, IA 52242, USA}
\author[0000-0002-1499-2667]{P.~Moriarty}\email{}\affiliation{School of Natural Sciences, University of Galway, University Road, Galway, H91 TK33, Ireland}
\author[0000-0002-3223-0754]{R.~Mukherjee}\email{}\affiliation{Department of Physics and Astronomy, Barnard College, Columbia University, NY 10027, USA}
\author[0000-0002-6121-3443]{W.~Ning}\email{}\affiliation{Department of Physics and Astronomy, University of California, Los Angeles, CA 90095, USA}
\author[0000-0002-4837-5253]{R.~A.~Ong}\email{}\affiliation{Department of Physics and Astronomy, University of California, Los Angeles, CA 90095, USA}
\author[0000-0001-7861-1707]{M.~Pohl}\email{}\affiliation{Institute of Physics and Astronomy, University of Potsdam, 14476 Potsdam-Golm, Germany and DESY, Platanenallee 6, 15738 Zeuthen, Germany}
\author[0000-0002-0529-1973]{E.~Pueschel}\email{}\affiliation{Fakult\"at f\"ur Physik \& Astronomie, Ruhr-Universit\"at Bochum, D-44780 Bochum, Germany}
\author[0000-0002-4855-2694]{J.~Quinn}\email{}\affiliation{School of Physics, University College Dublin, Belfield, Dublin 4, Ireland}
\author{P.~L.~Rabinowitz}\email{}\affiliation{Department of Physics, Washington University, St. Louis, MO 63130, USA}
\author[0000-0002-5351-3323]{K.~Ragan}\email{}\affiliation{Physics Department, McGill University, Montreal, QC H3A 2T8, Canada}
\author{P.~T.~Reynolds}\email{}\affiliation{Department of Physical Sciences, Munster Technological University, Bishopstown, Cork, T12 P928, Ireland}
\author[0000-0002-7523-7366]{D.~Ribeiro}\email{}\affiliation{School of Physics and Astronomy, University of Minnesota, Minneapolis, MN 55455, USA}
\author{E.~Roache}\email{}\affiliation{Center for Astrophysics $|$ Harvard \& Smithsonian, Cambridge, MA 02138, USA}
\author[0000-0003-1387-8915]{I.~Sadeh}\email{}\affiliation{DESY, Platanenallee 6, 15738 Zeuthen, Germany}
\author[0000-0002-3171-5039]{L.~Saha}\email{}\affiliation{Center for Astrophysics $|$ Harvard \& Smithsonian, Cambridge, MA 02138, USA}
\author{G.~H.~Sembroski}\email{}\affiliation{Department of Physics and Astronomy, Purdue University, West Lafayette, IN 47907, USA}
\author[0000-0002-9856-989X]{R.~Shang}\email{}\affiliation{Department of Physics and Astronomy, Barnard College, Columbia University, NY 10027, USA}
\author[0000-0002-9852-2469]{D.~Tak}\email{}\affiliation{SNU Astronomy Research Center, Seoul National University, Seoul 08826, Republic of Korea.}
\author{A.~K.~Talluri}\email{}\affiliation{School of Physics and Astronomy, University of Minnesota, Minneapolis, MN 55455, USA}
\author{J.~V.~Tucci}\email{}\affiliation{Department of Physics, Indiana University Indianapolis, Indianapolis, Indiana 46202, USA}
\author[0000-0002-8090-6528]{J.~Valverde}\email{}\affiliation{Department of Physics, University of Maryland, Baltimore County, Baltimore MD 21250, USA and NASA GSFC, Greenbelt, MD 20771, USA}
\author[0000-0003-2740-9714]{D.~A.~Williams}\email{}\affiliation{Santa Cruz Institute for Particle Physics and Department of Physics, University of California, Santa Cruz, CA 95064, USA}
\author[0000-0002-2730-2733]{S.~L.~Wong}\email{}\affiliation{Physics Department, McGill University, Montreal, QC H3A 2T8, Canada}
\author[0009-0001-6471-1405]{J.~Woo}\email{}\affiliation{Columbia Astrophysics Laboratory, Columbia University, New York, NY 10027, USA}

\collaboration{80}{(The VERITAS collaboration)}

\author{R.~Alfaro}
\affiliation{Instituto de F'{i}sica, Universidad Nacional Autónoma de México, Ciudad de Mexico, Mexico }

\author{C.~Alvarez}
\affiliation{Universidad Autónoma de Chiapas, Tuxtla Gutiérrez, Chiapas, México}

\author{J.C.~Arteaga-Velázquez}
\affiliation{Universidad Michoacana de San Nicolás de Hidalgo, Morelia, Mexico }

\author{D.~Avila Rojas}
\affiliation{Instituto de Astronom'{i}a, Universidad Nacional Autónoma de México, Ciudad de Mexico, Mexico }

\author{R.~Babu}
\affiliation{Department of Physics and Astronomy, Michigan State University, East Lansing, MI, USA }

\author{E.~Belmont-Moreno}
\affiliation{Instituto de F'{i}sica, Universidad Nacional Autónoma de México, Ciudad de Mexico, Mexico }

\author{A.~Bernal}
\affiliation{Instituto de Astronom'{i}a, Universidad Nacional Autónoma de México, Ciudad de Mexico, Mexico }

\author{K.S.~Caballero-Mora}
\affiliation{Universidad Autónoma de Chiapas, Tuxtla Gutiérrez, Chiapas, México}

\author{A.~Carramiñana}
\affiliation{Instituto Nacional de Astrof'{i}sica, Óptica y Electrónica, Puebla, Mexico }

\author{S.~Casanova}
\affiliation{Instytut Fizyki Jadrowej im Henryka Niewodniczanskiego Polskiej Akademii Nauk, IFJ-PAN, Krakow, Poland }

\author{U.~Cotti}
\affiliation{Universidad Michoacana de San Nicolás de Hidalgo, Morelia, Mexico }

\author{J.~Cotzomi}
\affiliation{Facultad de Ciencias F'{i}sico Matemáticas, Benemérita Universidad Autónoma de Puebla, Puebla, Mexico }

\author{E.~De la Fuente}
\affiliation{Departamento de F'{i}sica, Centro Universitario de Ciencias Exactase Ingenierias, Universidad de Guadalajara, Guadalajara, Mexico }

\author{C.~de León}
\affiliation{Universidad Michoacana de San Nicolás de Hidalgo, Morelia, Mexico }

\author{D.~Depaoli}
\affiliation{Max-Planck Institute for Nuclear Physics, 69117 Heidelberg, Germany}

\author{P.~Desiati}
\affiliation{Department of Physics, University of Wisconsin-Madison, Madison, WI, USA }

\author{N.~Di Lalla}
\affiliation{Department of Physics, Stanford University: Stanford, CA 94305–4060, USA}

\author{R.~Diaz Hernandez}
\affiliation{Instituto Nacional de Astrof'{i}sica, Óptica y Electrónica, Puebla, Mexico }

\author{M.A.~DuVernois}
\affiliation{Department of Physics, University of Wisconsin-Madison, Madison, WI, USA }

\author{K.~Engel}
\affiliation{Department of Physics, University of Maryland, College Park, MD, USA }

\author{T.~Ergin}
\affiliation{Department of Physics and Astronomy, Michigan State University, East Lansing, MI, USA }

\author{C.~Espinoza}
\affiliation{Instituto de F'{i}sica, Universidad Nacional Autónoma de México, Ciudad de Mexico, Mexico }

\author{K.L.~Fan}
\affiliation{Department of Physics, University of Maryland, College Park, MD, USA }

\author{N.~Fraija}
\affiliation{Instituto de Astronom'{i}a, Universidad Nacional Autónoma de México, Ciudad de Mexico, Mexico }

\author{S.~Fraija}
\affiliation{Instituto de Astronom'{i}a, Universidad Nacional Autónoma de México, Ciudad de Mexico, Mexico }

\author{J.A.~García-González}
\affiliation{Tecnologico de Monterrey, Escuela de Ingeniería y Ciencias, Ave. Eugenio Garza Sada 2501}

\author{F.~Garfias}
\affiliation{Instituto de Astronom'{i}a, Universidad Nacional Autónoma de México, Ciudad de Mexico, Mexico }

\author{A.~Gonzalez Muñoz}
\affiliation{Instituto de F'{i}sica, Universidad Nacional Autónoma de México, Ciudad de Mexico, Mexico }

\author{M.M.~González}
\affiliation{Instituto de Astronom'{i}a, Universidad Nacional Autónoma de México, Ciudad de Mexico, Mexico }

\author{J.A.~Goodman}
\affiliation{Department of Physics, University of Maryland, College Park, MD, USA }

\author{S.~Groetsch}
\affiliation{Department of Physics, Michigan Technological University, Houghton, MI, USA }

\author{J.P.~Harding}
\affiliation{Physics Division, Los Alamos National Laboratory, Los Alamos, NM, USA }

\author{S.~Hernández-Cadena}
\affiliation{Tsung-Dao Lee Institute, Shanghai Jiao Tong University, Shanghai, China}

\author{I.~Herzog}
\affiliation{Department of Physics and Astronomy, Michigan State University, East Lansing, MI, USA }

\author{D.~Huang}
\affiliation{Department of Physics, University of Maryland, College Park, MD, USA }

\author{F.~Hueyotl-Zahuantitla}
\affiliation{Universidad Autónoma de Chiapas, Tuxtla Gutiérrez, Chiapas, México}

\author{P.~Hüntemeyer}
\affiliation{Department of Physics, Michigan Technological University, Houghton, MI, USA }

\author{A.~Iriarte}
\affiliation{Instituto de Astronom'{i}a, Universidad Nacional Autónoma de México, Ciudad de Mexico, Mexico }

\author{S.~Kaufmann}
\affiliation{Universidad Politecnica de Pachuca, Pachuca, Hgo, Mexico }

\author{A.~Lara}
\affiliation{Instituto de Geof'{i}sica, Universidad Nacional Autónoma de México, Ciudad de Mexico, Mexico }

\author{J.~Lee}
\affiliation{University of Seoul, Seoul, Rep. of Korea}

\author{H.~León Vargas}
\affiliation{Instituto de F'{i}sica, Universidad Nacional Autónoma de México, Ciudad de Mexico, Mexico }

\author{A.L.~Longinotti}
\affiliation{Instituto de Astronom'{i}a, Universidad Nacional Autónoma de México, Ciudad de Mexico, Mexico }

\author{G.~Luis-Raya}
\affiliation{Universidad Politecnica de Pachuca, Pachuca, Hgo, Mexico }

\author{K.~Malone}
\affiliation{Physics Division, Los Alamos National Laboratory, Los Alamos, NM, USA }

\author{O.~Martinez}
\affiliation{Facultad de Ciencias F'{i}sico Matemáticas, Benemérita Universidad Autónoma de Puebla, Puebla, Mexico }

\author{J.~Martínez-Castro}
\affiliation{Centro de Investigaci'on en Computaci'on, Instituto Polit'ecnico Nacional, M'exico City, M'exico.}

\author{J.A.~Matthews}
\affiliation{Dept of Physics and Astronomy, University of New Mexico, Albuquerque, NM, USA }

\author{P.~Miranda-Romagnoli}
\affiliation{Universidad Autónoma del Estado de Hidalgo, Pachuca, Mexico }

\author{J.A.~Morales-Soto}
\affiliation{Universidad Michoacana de San Nicolás de Hidalgo, Morelia, Mexico }

\author{E.~Moreno}
\affiliation{Facultad de Ciencias F'{i}sico Matemáticas, Benemérita Universidad Autónoma de Puebla, Puebla, Mexico }

\author{M. ~Araya}
\affiliation{Universidad de Costa Rica}

\author{M.~Mostafá}
\affiliation{Department of Physics, Temple University, Philadelphia, PA, USA}

\author{M.~Najafi}
\affiliation{Department of Physics, Michigan Technological University, Houghton, MI, USA }

\author{A.~Nayerhoda}
\affiliation{Instytut Fizyki Jadrowej im Henryka Niewodniczanskiego Polskiej Akademii Nauk, IFJ-PAN, Krakow, Poland }

\author{L.~Nellen}
\affiliation{Instituto de Ciencias Nucleares, Universidad Nacional Autónoma de Mexico, Ciudad de Mexico, Mexico }

\author{N.~Omodei}
\affiliation{Department of Physics, Stanford University: Stanford, CA 94305–4060, USA}

\author{E.~Ponce}
\affiliation{Facultad de Ciencias F'{i}sico Matemáticas, Benemérita Universidad Autónoma de Puebla, Puebla, Mexico }

\author{E.G.~Pérez-Pérez}
\affiliation{Universidad Politecnica de Pachuca, Pachuca, Hgo, Mexico }

\author{C.D.~Rho}
\affiliation{Department of Physics, Sungkyunkwan University, Suwon 16419, South Korea}

\author{D.~Rosa-González}
\affiliation{Instituto Nacional de Astrof'{i}sica, Óptica y Electrónica, Puebla, Mexico }

\author{M.~Roth}
\affiliation{Los Alamos National Laboratory, Los Alamos, NM, USA}

\author{H.~Salazar}
\affiliation{Facultad de Ciencias F'{i}sico Matemáticas, Benemérita Universidad Autónoma de Puebla, Puebla, Mexico }

\author{A.~Sandoval}
\affiliation{Instituto de F'{i}sica, Universidad Nacional Autónoma de México, Ciudad de Mexico, Mexico }

\author{M.~Schneider}
\affiliation{Department of Physics, University of Maryland, College Park, MD, USA }

\author{J.~Serna-Franco}
\affiliation{Instituto de F'{i}sica, Universidad Nacional Autónoma de México, Ciudad de Mexico, Mexico }

\author{A.J.~Smith}
\affiliation{Department of Physics, University of Maryland, College Park, MD, USA }

\author{Y.~Son}
\affiliation{University of Seoul, Seoul, Rep. of Korea}

\author{R.W.~Springer}
\affiliation{Department of Physics and Astronomy, University of Utah, Salt Lake City, UT, USA }

\author{O.~Tibolla}
\affiliation{Universidad Politecnica de Pachuca, Pachuca, Hgo, Mexico }

\author{K.~Tollefson}
\affiliation{Department of Physics and Astronomy, Michigan State University, East Lansing, MI, USA }

\author{I.~Torres}
\affiliation{Instituto Nacional de Astrof'{i}sica, Óptica y Electrónica, Puebla, Mexico }

\author{R.~Torres-Escobedo}
\affiliation{SJTU}

\author{R.~Turner}
\affiliation{Department of Physics, Michigan Technological University, Houghton, MI, USA }

\author{F.~Ureña-Mena}
\affiliation{Instituto Nacional de Astrof'{i}sica, Óptica y Electrónica, Puebla, Mexico }

\author{E.~Varela}
\affiliation{Facultad de Ciencias F'{i}sico Matemáticas, Benemérita Universidad Autónoma de Puebla, Puebla, Mexico }

\author{L.~Villaseñor}
\affiliation{Facultad de Ciencias F'{i}sico Matemáticas, Benemérita Universidad Autónoma de Puebla, Puebla, Mexico }

\author{X.~Wang} \affiliation{Department of Physics, Michigan Technological University, Houghton, MI, USA} \affiliation{Department of Physics, Missouri University of Science and Technology, Rolla, MO, USA}


\author{Z.~Wang}
\affiliation{Department of Physics, University of Maryland, College Park, MD, USA }

\author{I.J.~Watson}
\affiliation{University of Seoul, Seoul, Rep. of Korea}

\author{H.~Wu}
\affiliation{Department of Physics, University of Wisconsin-Madison, Madison, WI, USA }

\author{S.~Yu}
\affiliation{Department of Physics, Pennsylvania State University, University Park, PA, USA }

\author{S.~Yun-Cárcamo}
\affiliation{Department of Physics, University of Maryland, College Park, MD, USA }

\author{H.~Zhou}
\affiliation{Tsung-Dao Lee Institute, Shanghai Jiao Tong University, Shanghai, China}

\author{M.~Martin}
\affiliation{Department of Physics and Astronomy, University of Utah, Salt Lake City, UT, USA }

\collaboration{100}{(The HAWC collaboration)}

\author[0000-0002-9709-5389]{Kaya Mori}
\affiliation{Columbia Astrophysics Laboratory, 550 West 120th Street, New York, NY 10027, USA}
\author[0000-0002-3681-145X]{Charles J. Hailey}
\affiliation{Columbia Astrophysics Laboratory, 550 West 120th Street, New York, NY 10027, USA}
\author[0000-0001-6189-7665]{Samar Safi-Harb}
\affiliation{Department of Physics and Astronomy, University of Manitoba, Winnipeg, MB R3T 2N2, Canada}
\author[0000-0002-2967-790X]{Shuo Zhang}
\affiliation{Department of Physics and Astronomy, Michigan State University, East Lansing, MI 48824, USA}
\collaboration{4}{(The \xmm{} collaboration)}









\begin{abstract}

We report observations of the ultra-high-energy gamma-ray source LHAASO J2108$+$5157, utilizing VERITAS, HAWC, \emph{Fermi}-LAT, and \xmm{}. VERITAS has collected $\sim$ 40 hours of data that we used to set ULs to the emission above 200 GeV. 
The HAWC data, collected over $\sim 2400$ days, reveal emission between 3 and 146 TeV, with a significance of $7.5~\sigma$, favoring an extended source model. The best-fit spectrum measured by HAWC is characterized by a simple power-law with a spectral index of $2.45\pm0.11_{stat}$. \fermilat{} analysis finds a point source with a very soft spectrum in the LHAASO J2108+5157 region, consistent with the 4FGL-DR3 catalog results. 
The \xmm{} analysis yields a null detection of the source in the 2 - 7 keV band. The broadband spectrum can be interpreted as a pulsar and a pulsar wind nebula system, where the GeV gamma-ray emission originates from an unidentified pulsar, and the X-ray and TeV emission is attributed to synchrotron radiation and inverse Compton scattering of electrons accelerated within a pulsar wind nebula. In this leptonic scenario, our X-ray upper limit provides a stringent constraint on the magnetic field, which is $\lesssim 1.5\ \mu$G.


\end{abstract}



\section{Introduction} \label{sec:intro}

It is generally recognized that cosmic rays (CRs) with energies up to $10^{15}$ eV, corresponding to the \emph{knee} in the CR particle spectrum, can be produced within our Galaxy \citep{Baade1934, Ginzburg1966, Blasi2013}. However, the locations and the nature of these powerful accelerators, often referred to as \emph{PeVatrons}, remain unknown. The charged nature of CRs make it difficult to trace their original direction, as their trajectories are significantly deflected by interactions with the Galactic magnetic field. However, in proximity to their source, CRs interact with matter or radiation fields, giving rise to gamma rays. Since gamma rays are neutral messengers unaffected by magnetic fields, we can trace their direction back to the point of origin.

Gamma rays, especially with ultra-high energy (UHE; $> 10^{14}$ eV), are a valuable tool for determining the characteristics of Galactic PeVatrons \citep{Bose2022}. In recent years, extensive air shower (EAS) arrays have provided evidence of gamma-ray emission above 100 TeV from a handful of objects in the Galactic Plane, including the Crab Nebula \citep{2019PhRvL.123e1101A}, eHWC J1825$-$134, eHWC J1907$+$063 and eHWC J2019$+$368 \citep{PhysRevLett.124.021102, kelly_icrc2023}. The list of these UHE emitters has been considerably extended by the results of the Large High Altitude Air Shower Observatory (LHAASO) experiment, a gamma-ray and CR observatory in the Chinese province of Sichuan \citep{2010ChPhC..34..249C}. The LHAASO collaboration reported the detection of 530 photons above 100 TeV and up to 1.4 PeV, from 12 Galactic sources. Each of these sources was detected with a statistical significance greater than $7~\sigma$ \citep{Cao_2021a}. The detection of gamma rays with energies close to 1 PeV from the Crab Nebula is the first model-independent evidence that this source is a leptonic PeVatron, and highlights the great discovery potential for such objects by this experiment.
LHAASO has recently expanded its catalog to around 90 sources, 43 of which were discovered above 100 TeV \citep{Lhaaso2023catalogue}. Among the TeV-PeV sources listed in \citet{Lhaaso2023catalogue}, there are 25 that have no counterpart in other wavelengths.

LHAASO J2108+5157 was detected by the LHAASO collaboration in the energy range from 1 to 25 TeV at $8.1~\sigma$ using the Water Cherenkov Detector Array (WCDA) and above 25 TeV at $30.3~\sigma$ with the Kilometer Squared Array (KM2A) detector \citep{Lhaaso2021catalogue, Cao_2021b, Lhaaso2023catalogue}. Despite its unambiguous detection in the TeV gamma-ray band, it is an interesting candidate for further investigation as it has not been detected at any other wavelength. The power-law index of its spectrum is reported to change from $1.56 \pm 0.34$ in the 1-25 TeV range to $2.97 \pm 0.07$ above 25 TeV. Initially, LHAASO J2108+5157 was identified as a point source using only the KM2A detector \citep{Cao_2021b}. However, with a larger dataset from KM2A and the inclusion of the WCDA detector, the source is reported to be slightly extended and can be modeled as 2D-Gaussian, with sigma values of $0.19^{\circ} \pm 0.02^{\circ}$ and $0.14^{\circ} \pm 0.03^{\circ}$ in the KM2A and WCDA data, respectively \citep{Lhaaso2023catalogue}. 

The Large-Sized Telescope prototype (LST-1) observed this source for 49 hours; no detection was reported \citep{Abe23}. 
In a dedicated analysis of the region around LHAASO J2108+5157 using 12.2 years of \fermilat data, a hard-spectrum source was detected at the $4~\sigma$ level, with a photon index of $1.9 \pm 0.2$ \citep{Abe23}, in addition to the previously identified soft-spectrum source 4FGL J2108+5155, which shows no emission above 2 GeV \citep{Cao_2021b}. The steep spectrum of 4FGL J2108+5155 above a few GeV makes it incompatible with the LHAASO spectral measurement. The hard-spectrum source has an angular separation of $\sim 0.27^{\circ}$ from LHAASO J2108+5157. Since this distance is greater than the extension upper limit (UL) reported in \citet{Cao_2021b}, it is unlikely that this hard spectrum source is associated with LHAASO J2108+5157. 

No significant X-ray emission was detected during the 4.7 ks exposure from the Swift-XRT survey of the LHAASO J2108+5157 region \citep{Stroh_2013}. The nearest known X-ray source is the binary RX J2107.3$+$5202, located $0.25^{\circ}$ from LHAASO J2108+5157. Moreover, no energetic pulsar has been identified in the nearby region.

Despite recent progress in the discovery of PeV gamma-ray sources, in particular with HAWC and LHAASO, the fundamental question of the nature of objects producing gamma-rays above 100 TeV remains unanswered. 
With their ability to resolve UHE sources with an angular resolution $\le 0.1^{\circ}$, the imaging atmospheric Cherenkov telescopes (IACTs) offer a complementary perspective that provides more insight into the identification of gamma-ray sources. In addition, precise spectral measurements in the GeV-TeV range can help distinguish between leptonic and hadronic \emph{PeVatrons}, as the Klein–Nishina suppression renders inverse Compton scattering of non-thermal electrons inefficient, leading to suppression of the leptonic emission channel in the TeV range.

The structure of the paper is as follows: Sections 2,3,4 and 5 describe the data analysis procedures and the results for VERITAS, HAWC, \fermilat{}, and \xmm{}, respectively. In Section 6, we discuss the multi-wavelength modelling of LHAASO J2108+5158, and the results of this study are summarized in the concluding Section 7.

\section{VERITAS observations and data analysis} \label{sec:VERITAS_analysis}

The Very Energetic Radiation Imaging Telescope Array System (VERITAS) is an array of four IACTs located at the Fred Lawrence Whipple Observatory in Amado, Arizona \citep{2006APh....25..391H}. Each telescope is equipped with a 12-meter tessellated reflector and a 499-element photomultiplier tube (PMT) camera, providing a field of view (FoV) of $3.5^\circ$. These telescopes capture the Cherenkov light produced by gamma-ray and cosmic-ray showers in the atmosphere. VERITAS is sensitive in the energy range between $\sim$80 GeV and $\sim$30 TeV. It has an angular resolution of $\sim0.1^{\circ}$ at 1 TeV, and can detect a point source with 1\% of the flux of the Crab Nebula, with $5~\sigma$ statistical significance, in 24 hours\footnote{\url{https://veritas.sao.arizona.edu/about-veritas/veritas-specifications}}.

\begin{figure}[!h]
\begin{center} 
  \includegraphics[width=0.45\textwidth]{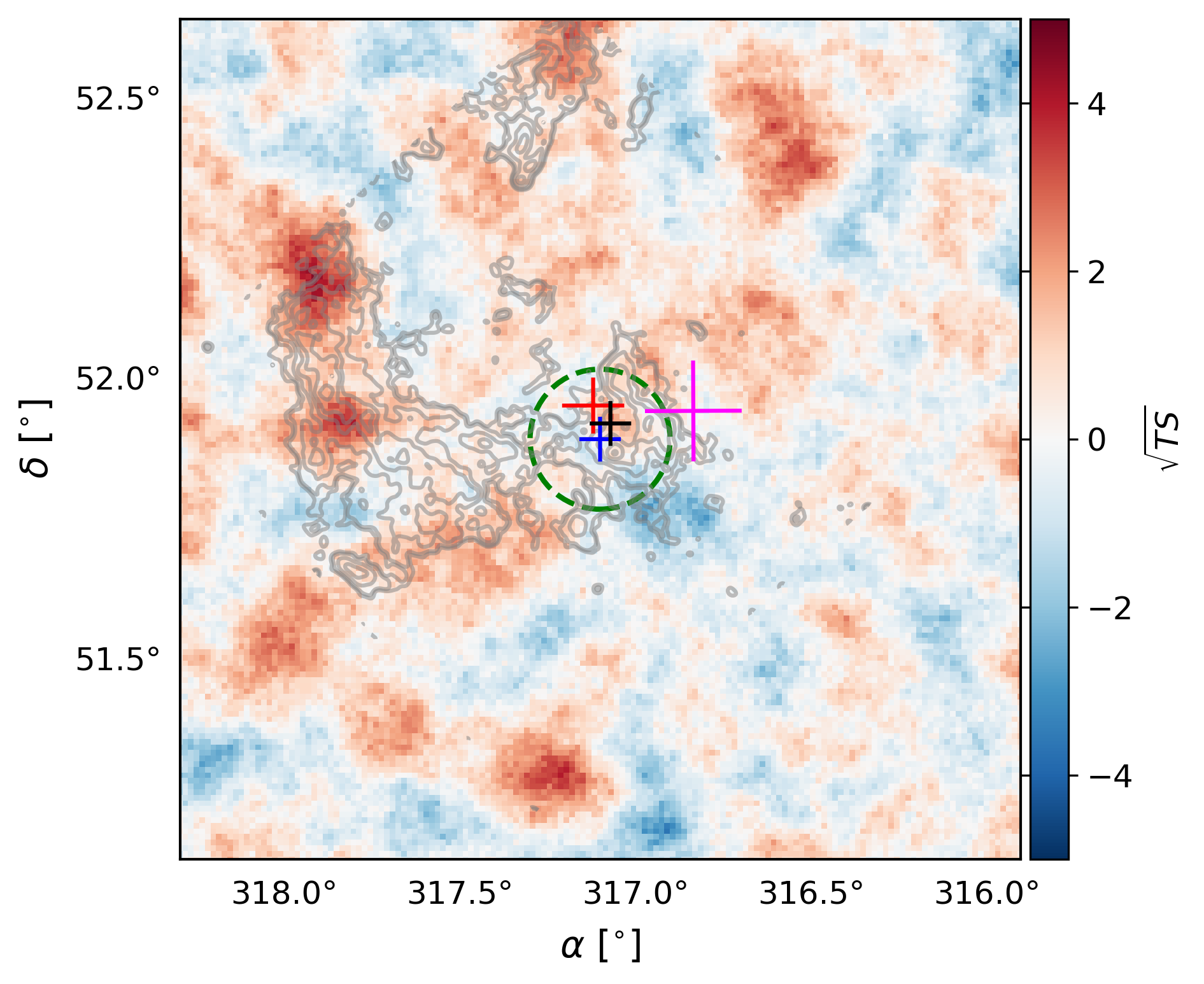}
  \caption{VERITAS significance map above 200 GeV, created with an integration radius of $0.09^{\circ}$. The green dotted circle shows the integration radius ($\theta=0.25^{\circ}$) used to extract spectral ULs from VERITAS. The best-fit positions measured by WCDA, KM2A, HAWC and \fermilat with their error bars are also shown as magenta, blue, red and black plus signs respectively. Light gray contours from the $^{13}$CO map, integrated between -20 to -8 $\mathrm{km \ s^{-1}}$, are also shown. These contours correspond levels of [$-4$, 4, 5, 8, 12, 16, 20] times the rms value of $0.5 \ \mathrm{K \ km \ s^{-1}}$ \citep{Fuente2023II}.}
  \label{fig:2108_skymap_point}
\end{center}
\end{figure}

VERITAS observed LHAASO J2108+5157 for 40 hours in 2021. After applying quality cuts and a correction for dead time, we obtain 35 hours of good quality data. The observations were performed in wobble mode with an offset of $0.7^{\circ}$ to the source centroid (RA: $317.15^{\circ}$, Dec: $51.95^{\circ}$). 
A minimum of two images is required for event reconstruction, and a machine-learning classification method utilizing boosted decision trees \citep{Maria2017} was employed to remove background events. Despite removing more than $99\%$ of background events, there is still an irreducible background, estimated using the ring background method \citep{Berge2007}. The reconstruction and event selection led to an energy threshold of $200 \ \mathrm{GeV}$ for the analysis.

Figure~\ref{fig:2108_skymap_point} shows the significance map of the LHAASO J2108+5157 region using VERITAS data above 200 GeV. The map is smoothed with a circular window of radius $0.09^{\circ}$, consistent with the VERITAS point spread function (PSF). It is clear from the map that no gamma-ray excess is detected at the location of LHAASO J2108+5157 (RA = $317.15^{\circ}$, Dec = $51.95^{\circ}$). The statistical significance is calculated using the likelihood method \citep{LiMa1983}, resulting in a value of $0.6~\sigma$. The significance is also calculated by assuming LHAASO J2108+5157 as an extended source with a radius of $\theta=0.25^{\circ}$. This also leads to null detection at the $0.3~\sigma$ level. 
The spectral analysis is performed for a circular region with a radius of $0.25^{\circ}$ around the LHAASO J2108+5157-KM2A position since the source is detected as an extended source in \citet{Lhaaso2023catalogue}. The resulting ULs for the flux at a $95\%$ confidence level are shown in Figure~\ref{fig:joint-spectrum-fit}. These ULs are calculated for energies above a threshold of 500 GeV. 




\section{HAWC analysis} \label{sec:HAWC_analysis}
The High Altitude Water Cherenkov Gamma-Ray Observatory (HAWC) is a ground-based water Cherenkov instrument located in Sierra Negra, Puebla state, Mexico, at an altitude of 4,100 meters above sea level \citep{NIM2023}. It consists of 300 tanks in the main array. Each tank is equipped with three 8-inch Hamamatsu PMTs and one 10-inch high-quantum efficiency Hamamatsu PMT. HAWC is sensitive to extensive air shower (EAS) events with primary energies from several hundreds of GeV to above 100~TeV. It has a duty cycle of more than 95~\%. 

\begin{figure}[!b]
\begin{center} 
  \includegraphics[width=0.45\textwidth]{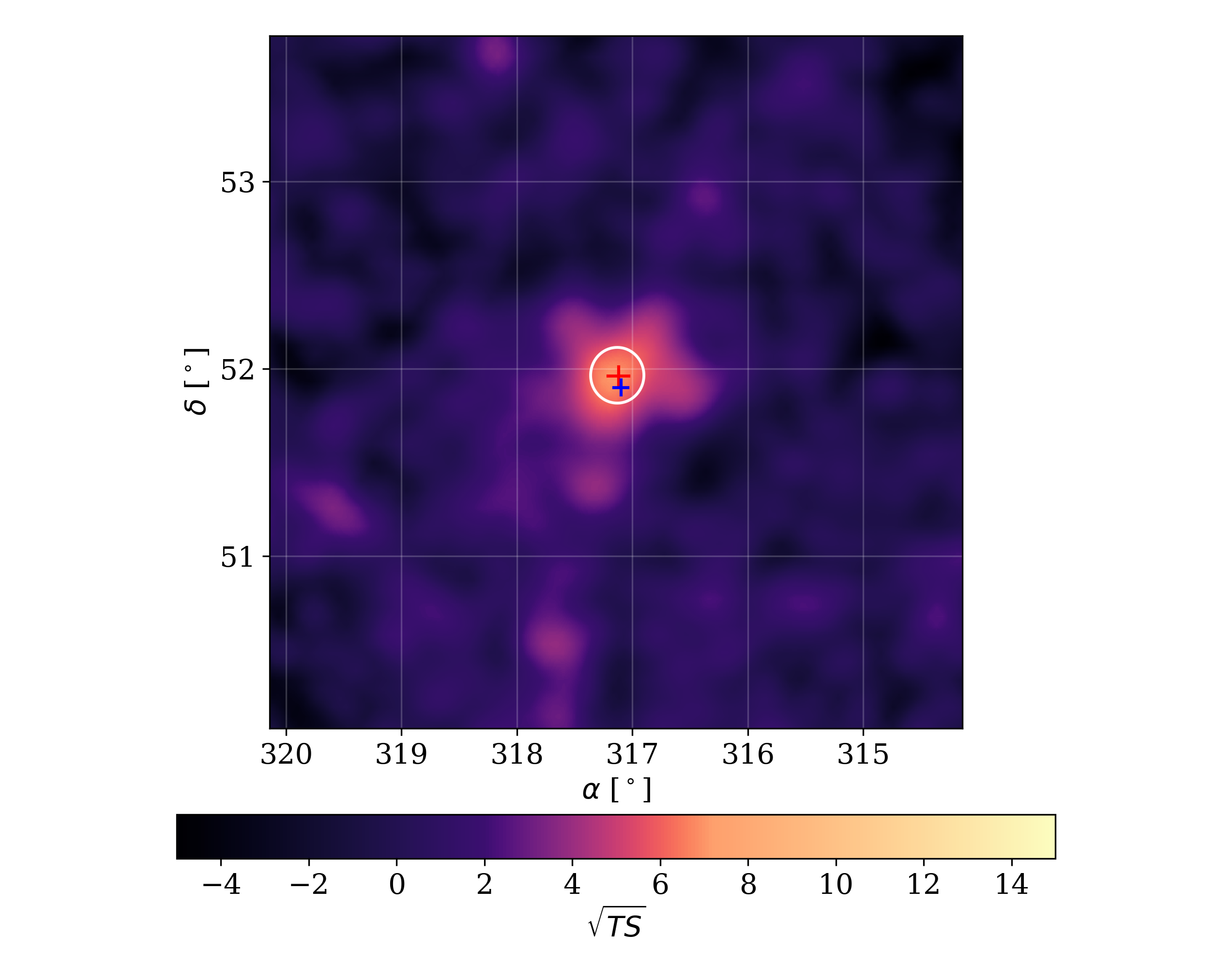}
  \caption{HAWC significance map of the LHAASO J2108+5157 region above 300~GeV. The best-fit positions measured by HAWC and KM2A are represented as red and blue markers, respectively, along with their error bars. A white circle around the HAWC position indicates the best-fit extension.}
  \label{fig:2108_sigmap}
\end{center}
\end{figure}

Using the newly produced pass-5 data ($\sim $2400~days) \citep{Pass5Hawc2024}, a source in the LHAASO J2108+5157 region was detected significantly above 300~GeV,
as shown in Figure~\ref{fig:2108_sigmap}. The HAWC data are binned into two-dimensional bins based on the fraction of PMT hits and the reconstructed energy.
Since events with shower cores landing on the detector array (so-called ``on-array" events) allow a more accurate reconstruction, only the ``on-array" events were used in this analysis.
The morphology and energy spectrum of the source is determined by likelihood fitting with the HAWC Accelerated Likelihood (HAL) plugin for the Multi-Mission Maximum Likelihood (3ML) framework~\citep{vianello2015multi}. Likelihood calculation with the HAL plugin is well described in previous HAWC publications (e.g., \citep{younk2015high,hawc2017crab, Abeysekara:2021DF}). 

\begin{figure}[!h]
\begin{center} 
  \includegraphics[width=0.45\textwidth]{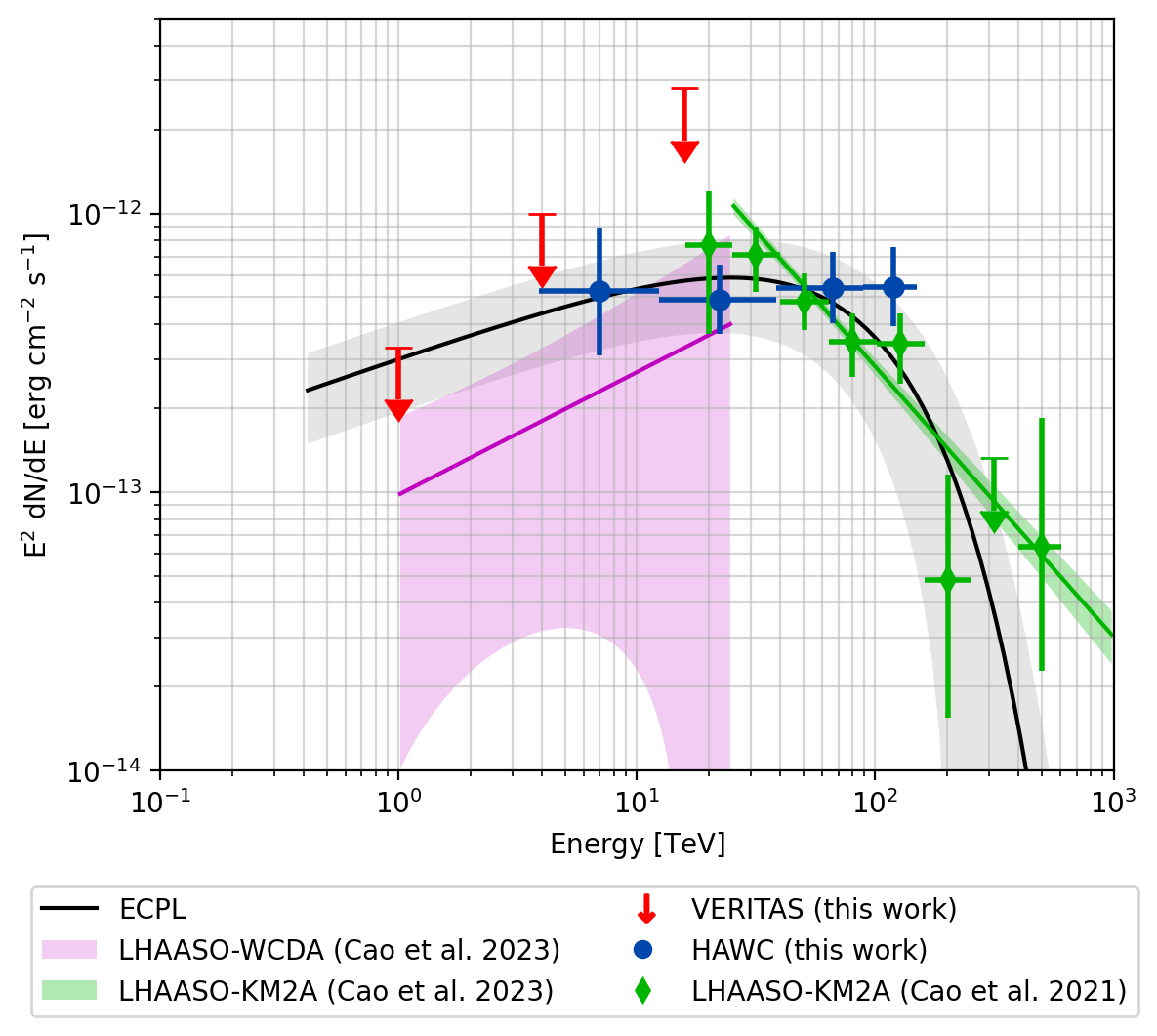}
  \caption{The spectral energy distribution of \lhaaso{} measured by VERITAS and HAWC. Notably, the VERITAS ULs are derived from a circular region with $0.25^{\circ}$ radius for energies above a threshold of 500 GeV. The data from LHAASO-KM2A reported in \citet{Cao_2021b} were included in the joint fit. The fitting model used is a power-law with exponential cutoff (ECPL). The spectral index is fixed to a value of 1.7, consistent with \citet{Lhaaso2023catalogue}. This fitting constrains the cutoff energy to $82\pm30 \ \mathrm{TeV}$. The gray band represents the $68\%$ error band of the ECPL model. The magenta and green bands represent the power-law spectral energy distribution plots for WCDA and KM2A, respectively, as reported in \citet{Lhaaso2023catalogue}.}
  \label{fig:joint-spectrum-fit}
\end{center}
\end{figure}

We chose a circular region of interest (RoI) with a radius of $3^{\circ}$ centered at RA=$317.14^{\circ}$, Dec=$51.94^{\circ}$ Since the source is about $3^{\circ}$ away from the Galactic plane, we assume that the diffuse background emission from the Galactic diffuse emission and unresolved sources is negligible in this analysis \citep{HAWC_geminga_2017, hess2014diffuse}. Two morphology models were tested with the HAWC data, and the extended source model with a symmetric Gaussian was favored. It yielded a detection significance of $7.5~\sigma$, with the best-fit centroid location at RA=$317.12^{\circ} \pm 0.09^{\circ}$ and Dec=$51.96^{\circ} \pm 0.05^{\circ}$, and a best-fit extension of $0.21^{\circ} \pm 0.04^{\circ}$ (see Figure~\ref{fig:2108_sigmap}). The energy spectrum is well fit by a power law ($dN/dE = N_{0} (E/E_{0})^{-\Gamma}$) with flux normalization of 
$ 1.86 _{-0.30}^ {+0.40} (stat.)^{+0.24}_{-0.17} (syst.)\times$ $10^{-16}$ $\rm cm^{-2} TeV^{-1} s^{-1}$ at a pivot energy of 35~TeV and an index of $2.45 \pm 0.11 (stat.)^{+0.01}_{-0.03} (syst.)$. 
After determining the best-fit spectral parameters from the likelihood fitting across the whole energy range, the flux points are calculated by re-fitting the flux normalization in each energy bin while fixing the spectral index to the global best-fit value. The resulting fluxes correspond to the median energy of each bin (see Figure~\ref{fig:joint-spectrum-fit} for HAWC spectral points). The energy range of 3 -- 146 TeV is determined by applying a step function to the best-fit spectral model and varying its boundaries to find where the log-likelihood significantly deviates from the best fit~\citep{abeysekara2018ss433}. This range covers the transition between the spectra measured by LHAASO-WCDA (index of –1.56 $\pm$ 0.34) and KM2A (index of –2.83 $\pm$ 0.18~\citep{Lhaaso2023catalogue}), and motivates the search for a spectral curvature in the HAWC data. With the current statistics, the HAWC data do not support spectral curvature. A log-parabola model 
($dN/dE = N_{0} (E/E_{0})^{-\alpha - \beta \ln (E/E_{0})})$
is disfavored relative to a power law model based on the Bayesian information criterion (BIC), with a $\Delta$BIC of 11 \citep{Kass_1995}. Nevertheless, the HAWC results are consistent with LHAASO within uncertainties and helps bridge the spectra observed by WCDA and KM2A. A similar test with increased statistics in the future could lead to a more definitive conclusion. Current best-fit parameters for a log-parabola model are a flux normalization $ N_{0} =  2.3 _{-0.5}^ {+0.6}\times$ $10^{-16}$ 
$\rm cm^{-2} TeV^{-1} s^{-1}$,   $\alpha = 2.52\pm$0.21, and $\beta = 0.17 \pm 0.18$. 

The systematic errors were calculated as described in \citet{hawc2019crab}, in which different sources of systematic uncertainty are investigated, including the charge uncertainty, PMT threshold, late-light simulation, and absolute PMT efficiency/time dependence. They are treated separately and the effects of each source of systematic uncertainty are added in quadrature to the others to obtain the final systematic errors for the source. 

We also performed a joint fit using flux points from VERITAS, HAWC, and LHAASO. Among these independent measurements, HAWC provides the strongest constraints below 20 TeV, as shown in Figure~\ref{fig:joint-spectrum-fit}. For this fit, we used a power law with an exponential cutoff (ECPL), described by the equation ($dN/dE = N_{0} (E/E_{0})^{-\gamma} exp (-E/E_{cutoff}$). The spectral index ($\gamma$) and reference energy ($E_{0}$) are fixed at a value of 1.7 and 20 TeV respectively, resulting in a cutoff energy of  $82\pm30$ TeV. The spectral index is fixed at 1.7 to ensure consistency with the VERITAS ULs and to better constrain the cutoff energy.



\section{\emph{Fermi}-LAT analysis} \label{sec:Fermi_analysis}

To investigate any GeV emission associated with \lhaaso{}, we analyzed 14.2 years of the Large Area Telescope (LAT) data (August 2008 - October 2022, MET 239557417 - 687054166). The LAT is a pair-conversion detector on board the Fermi Gamma-ray Space Telescope (\emph{Fermi}). 
It can detect gamma rays in the energy range from below 20 MeV to above 300 GeV with an energy-dependent angular resolution of $\lesssim$ $0.2^{\circ}$ above 10 GeV \citep{2009ApJ...697.1071A}. 

The region of interest (RoI) was defined as a box region with a side length of $21^{\circ}$ (acceptance cone radius of $15^{\circ}$) centered at the centroid of \lhaaso{} (RA=$317.22^{\circ}$, Dec=$51.95^{\circ}$). We selected the ``SOURCE" class (evclass = 128) and ``FRONT\&BACK" type (evtype = 3) events in 100 MeV -- 1 TeV within the RoI. The events were reconstructed using the instrument response function (IRF) \texttt{P8R3\_SOURCE\_V3}. We filtered the events with a maximum zenith angle of 90$^{\circ}$ and the filter expression \texttt{DATA\_QUAL>0 \&\& LAT\_CONFIG==1}. We then performed a binned likelihood analysis using \texttt{Fermipy v1.2} \citep{2017ICRC...35..824W}, a Python package for analyzing the LAT data with the Fermi Science Tools. The events were binned into $0.1^{\circ}$ spatial bins and 8 logarithmic energy bins per decade. Our model comprised the Galactic diffuse emission model (\texttt{gll\_iem\_v07.fits}), the isotropic emission model (\texttt{iso\_P8R3\_SOURCE\_V3\_v1.txt}), and the source models within $20^{\circ}$ of the center of the ROI from the latest LAT source catalog (4FGL-DR3; \citet{Abdollahi_2022}). The model was fitted to the data to obtain the maximum likelihood. 

The data are well-explained by the model and Gaussian fluctuations. We have generated and analyzed test statistic (TS\footnote{TS $=-2$ln$(L_{max,0}/L_{max,1})$ was calculated for each spatial bin, where in each spatial bin $L_{max,0}$ and $L_{max,1}$ are the maximum likelihoods without and with an additional source, respectively. For the spectrum of an additional source, a power law with index 2 was used.}) maps in different energy ranges (1, 3, 5, 10, 30, 50, and 100 GeV -- 1 TeV) to search for any significant gamma-ray excess in the vicinity of \lhaaso; no significant excess is detected. \fermi\ is the only source in 4FGL-DR3 that lies within the extension of \lhaaso\ ($0.19^{\circ} \pm 0.02^{\circ}$). As a point source that is $0.13^{\circ}$ away from \lhaaso, \fermi\ is modeled with a log parabola in 4FGL-DR3. The parameters of our best-fit model for \fermi\ are in good agreement with those of 4FGL-DR3. \fermi{} is detected below 10 GeV and shows a sharp cutoff around 1 GeV in its spectrum. Such a spectral feature is often observed in gamma-ray pulsars.

\cite{Cao_2021b} reported a significant ($7.8~\sigma$) detection of the spatial extension $\sim0.48^{\circ}$ (2-D Gaussian width) of \fermi{} using 12.2 years of the LAT data in the 1 GeV -- 1 TeV range. We performed an extension fit with a symmetric Gaussian model in this energy range, starting from our best-fit model and freeing the same parameters as before. The best-fit extension of \fermi{} is $0.55^{\circ}$ with a marginal increase in likelihood ($\Delta$ln$\mathcal{L}=16$) and lower statistical significance ($5.7~\sigma$). In addition, the significance of multiple sources located near \fermi{} dropped below the detection threshold after the model was re-optimized with \fermi{} as an extended source. It is likely that the gamma rays originally attributed to nearby point sources are now attributed to \fermi{} because the source model has been so greatly extended. The model parameters of the extended \fermi{} are poorly constrained. \citet{Abe23} used 13.5 years of LAT data in the 1--500 GeV band to claim evidence ($4~\sigma$) of a new hard (power-law differential index = 1.9) point source just outside the extent of \lhaaso{}. Despite the low possibility as the counterpart of \lhaaso{} due to its location, adding the new source to their model improved their model for \fermi{}. We do not detect this point source. Additionally, all the model parameters of \fermi\ are already tightly constrained in our model, with values that match the catalog. We attribute the discrepancy between our work and the previous work of \citet{Cao_2021b} and \citet{Abe23} to the energy range of the data used in each analysis. While \citet{Cao_2021b} and \citet{Abe23} fitted their model to the high-energy data ($>$ 1 GeV), our model was first optimized in the entire energy range of 4FGL-DR3 (100 MeV -- 1 TeV). Since the majority of the sources in the catalog, including \fermi, have most of their emission below a few GeV, excluding the low-energy data to fit the catalog models would naturally lead to deviations from the model. We conclude that the GeV emission in the \lhaaso{} region is characterized by a point-like source with a spectral cutoff at $\sim1$ GeV, consistent with the model for \fermi{} in 4FGL-DR3.

\section{\xmm{} analysis} \label{sec:XMM_analysis}
\xmm{} is an X-ray space observatory with three co-aligned telescopes on board. The primary instruments of \xmm{} are the European Photon Imaging Cameras (EPIC), which consists of three CCD cameras (MOS1, MOS2 and pn). Each camera covers a field of view with a diameter $0.5^{\circ}$ with an angular resolution of 6$\arcsec$ FWHM. The cameras are sensitive in the energy range from 0.2 to 12 keV (for MOS1 and MOS2, 15 keV for pn; \cite{xmmmos,xmmpn}).

We obtained new \xmm{} observations of \lhaaso{} in May 2023 (observation ID 0923400501, 0923400901, and 0923401001, total exposure 96 ks). The telescope pointing was at the centroid of \lhaaso{} reported in \cite{Cao_2021b}. All three cameras were operated in full frame mode with a thin filter. We utilized nearly the entire FoV (a circular region of radius $0.2^{\circ}$) for the analysis. Since a significant part of the source region lies on MOS 1's missing chips (CCD3 and CCD6), only MOS 2 and pn were used.

We processed the \xmm{} data using the XMM-Newton Extended Source Analysis Software (XMM-ESAS) package in the XMM-Newton Science Analysis System (SAS v21.0.0, \cite{SAS}). First, we visually examined the images from the three cameras. We created event files for MOS2 (pn) using the \textit{emchain} (\textit{epchain}) task and filtered the good time intervals (GTIs) affected by soft proton (SP) flares using the \textit{espfilt} task. The net exposure after filtering is 62 ks.

We used the \textit{cheese} task to detect and mask point sources in the FoV. With the exception of V1061 Cygni, an eclipsing binary located just outside our source region at the northwestern edge of the FoV, no bright point source is present. We generated model quiescent particle background (QPB) spectra and images from the corner chip data and the filter-wheel closed data using \textit{mosspectra} (\textit{pnspectra}) and \textit{mosback} (\textit{pnback}) tasks. We subtracted the QPB, merged the three observations into a mosaic and corrected the exposure to obtain an image of the FoV in the 2--7 keV range. This energy range was selected to avoid additional modeling of the thermal cosmic background (local bubble and halo) and the line emissions of the QBP (not included in the model) as well as the charge exchange in the solar wind. No significant emission was observed in the image in the vicinity of \lhaaso{}. 

We have performed a spectral analysis with \texttt{Xspec} \citep{xspec} to place a UL on a putative diffuse emission associated with \lhaaso{}. The spectra from the three observations and two detectors were jointly analyzed in the 2--7 keV range. The model QPB spectra were used for the background spectra while additional background components were included in the source model. The background components included in the source model are the weak line emission of the QPB (Cr K$\alpha$ at 5.4 keV and Fe K$\alpha$ at 6.4 keV), and the cosmic X-ray background (power law with $\Gamma=1.4$ and normalization 11.6 photons/keV/s/cm$^2$/sr, \citep{CXB}). The Galactic hydrogen column density in the direction of \lhaaso{} (N$_{H}=1.21\times10^{22}$ cm$^{-2}$) was used to account for the absorption\footnote{\url{https://www.swift.ac.uk/analysis/nhtot/index.php}}. Using an absorbed power law (\texttt{tbabs*pow}) with the power-law index fixed to 2, we calculate the 95\% unabsorbed flux UL in 2--10 keV $=3.5\times10^{-13}$ erg s$^{-1}$ cm$^{-2}$ (reduced $\chi^2=1922/2464$) for the X-ray emission from \lhaaso{}\footnote{When the power-law index is fixed to 1.5 (2.5), the UL is increased (decreased) by $1.2\times10^{-13}$ erg s$^{-1}$ cm$^{-2}$.}. Note that the residual SP background was not included in the model because the model parameters are not constrained due to the limited counts over the background and the degeneracy with the source model (power law). The effect of the residual SPs should be marginal -- we examined the count rate over the exposure and confirmed that none of the remaining GTIs had significantly elevated rates after filtering.


\section{Leptonic modeling of multiwavelength emission} \label{sec:Naima_modeling}
Gamma rays can be produced either by the leptonic scenario, in which relativistic electrons emit inverse Compton radiation by up-scattering low-energy photons, or by the hadronic scenario, in which relativistic protons interact with protons in the surrounding gas, leading to the production of neutral and charged pions. Since no strong pulsars or supernova remnants (SNRs) have yet been detected within $99\%$ containment radius of KM2A extension ($0.58^{\circ}$) of LHAASO J2108+5157, it is difficult to conclude anything firmly on the origin of gamma-ray emission. Recently, a hadronic scenario has been proposed in which cosmic rays escaping from an old SNR and interacting with nearby molecular clouds produce the gamma rays \citep{Sarkar2023, Fuente2023II, Alison2023}. Alternatively, the pulsar-like spectral signature of 4FGL J2108+5155 makes the leptonic scenario plausible to explain the gamma-ray emission from LHAASO J2108+5157 \citep{Cao_2021b, Abe23}. 

In this section, we provide a benchmark model for the leptonic scenario, which explains the multiwavelength data obtained as part of this work and from \cite{Cao_2021b} and \cite{Lhaaso2023catalogue}. Investigation of the broader parameter space is left for future work.

The VERITAS UL at 1 TeV, as well as the \fermilat{} UL at 200 GeV, shows a significant hardening of the spectrum below 10 TeV, which is critical for considering not only the leptonic but also the hadronic scenario where the accelerator (middle-aged SNR) and the gamma-ray emitter (dense gas cloud) are separated. A detailed study of this scenario, utilizing the HAWC, VERITAS, and \xmm{} data from this work and newly obtained radio data, is performed in de La Fuente et al. (in preparation).

Previous studies have already modeled the broadband gamma-ray emission by assuming an exponential cutoff power-law electron population with a spectral index of 2.2, a cutoff energy of 200 TeV, and a magnetic field of $3 \ \mathrm{\mu G}$ \citep{Cao_2021b}. However, with the inclusion of ULs from LST-1 data in the GeV-TeV range, the electron spectral index was constrained to a value of ($1.5\pm0.4$). Furthermore, by adding ULs in the X-ray band from \xmm{}, the maximum magnetic field is estimated to be $1.2\ \mathrm{\mu G}$ in the leptonic model \citep{Abe23}. 

In this new study, we include additional data from the VERITAS and HAWC observatories and re-examine the modeling under the leptonic scenario. We assume that \fermi{} is a pulsar that powers a pulsar wind nebula (PWN) associated with \lhaaso{}. Electrons accelerated within the PWN follow a power-law distribution with an exponential cutoff 
\begin{equation}
    dN_{e}/dE_{e} = N_{e,0}(E_{e}/E_{e,0})^{-\alpha_{e}}\exp{\left[-\left(\frac{E_e}{E_{e,cut}}\right)^{\beta}\right]},
\end{equation}
where $N_{e,0}$ is the normalization constant, representing the electron flux at the reference energy of $E_{e,0}$=1 TeV, $\alpha_{e}$ is the electron spectral index, and $\beta$ is the cutoff index whose value can be 1 (simple exponential cutoff) or 2 (super-exponential cutoff) depending on the particle acceleration mechanism \citep{ZA2007}.
The photons of the cosmic microwave background (CMB) are considered as seed photons with which relativistic electrons interact to produce emission in the VHE region through the process of inverse Compton scattering. The same electron population also generates non-thermal X-rays through the synchrotron process. The normalization constant $N_{e,0}$, and hence the total electron energy, is scaled to the estimated distance to the source (1 kpc, \cite{Cao_2021b}). The TeV and X-ray data are modeled using the \textit{Naima} package \citep{Zabalza2015}, while the \fermilat{} data are modeled using a log parabola spectrum \citep{Abdollahi_2022} with the best-fit parameters from this work.

\begin{figure}
\begin{center} 
  \includegraphics[width=0.45\textwidth]{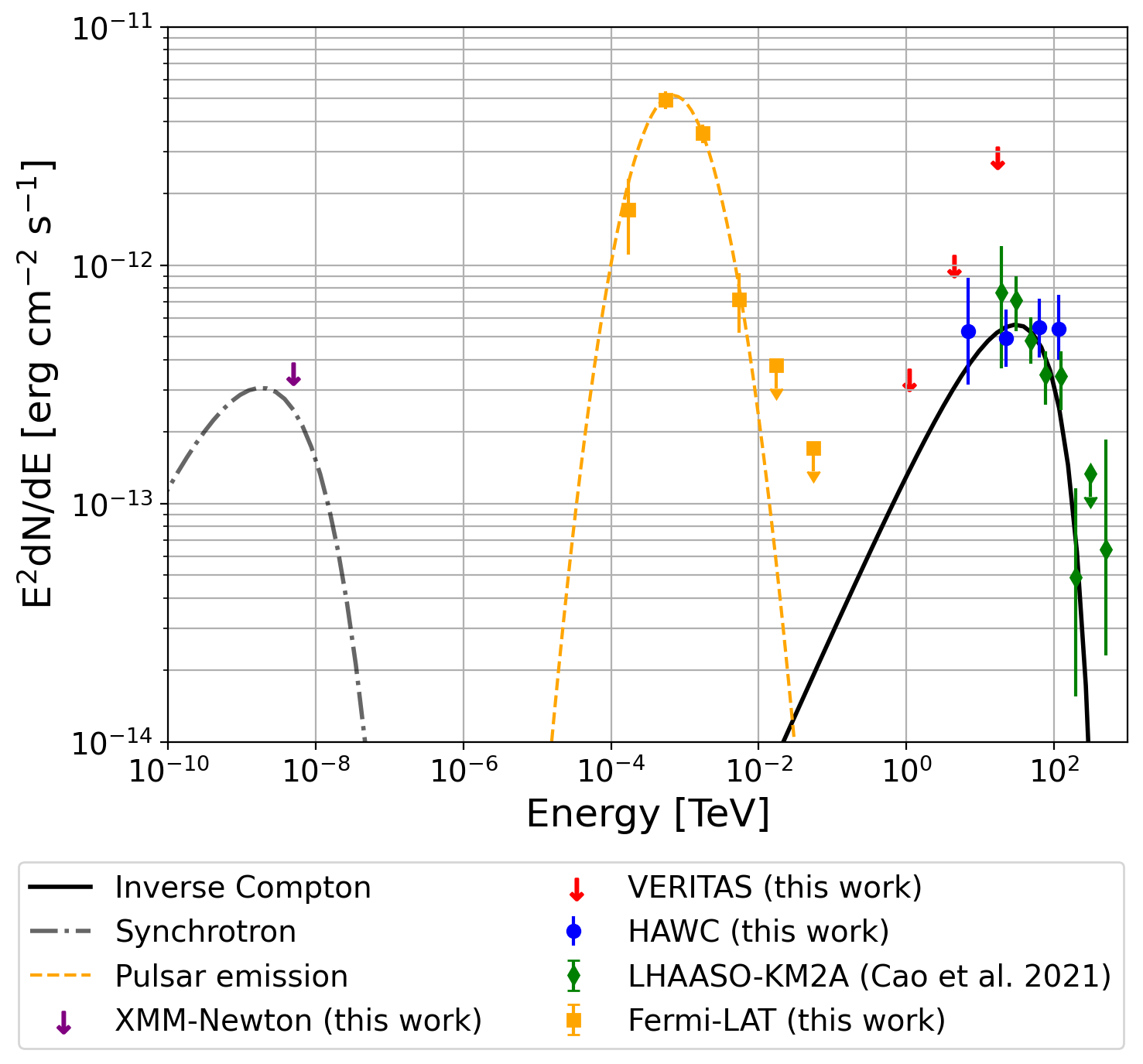}
  \caption{Leptonic model under the assumption of PWN as source class. The sharp cutoff around 100 TeV observed by LHAASO-KM2A favors a super-exponential cutoff ($\beta=2$) in the electron spectrum. Our benchmark model with the electron power-law index $\alpha_e=1.6$ and cutoff energy $E_{e,cut}=196$ TeV explains the VERITAS, HAWC, and LHAASO observations. Our \xmm{} flux UL constrains the magnetic field to $\lesssim1.5\ \mu$G.}
  \label{fig:pure_leptonic_model}
\end{center}
\end{figure}

The sharp cutoff in the 100s TeV range measured by LHAASO-KM2A favors a super-exponential cutoff ($\beta=2$) in the electron spectrum over a simple exponential cutoff ($\beta=1$). With $\beta$ fixed at 2, our benchmark model with $\alpha_e=1.6$, $E_{e,cut}=196$ TeV explains the observed TeV data well, as shown in Figure~\ref{fig:pure_leptonic_model}. A magnetic field of $\sim1.5\ \mu$G or below is required to ensure compliance with the X-ray flux UL. The total energy of electrons above 1 GeV is estimated to be $5.2 \times 10^{44} (\frac{d}{1.0\ \mathrm{kpc}})^{2}$ erg.

\section{Summary and conclusion}
In this paper, we have analyzed $62 \ \mathrm{ks}$ of XMM-Newton data, 14 years of \fermilat data, 40 hours of VERITAS data, and 2400 days of HAWC data. From the multi-wavelength analysis we draw the following conclusions:
\begin{itemize}
    \item The \xmm{} observation in the energy range 2-7 keV did not detect any significant signal from LHASSO J2108+5157. Despite this null detection, the ULs obtained in the X-ray region allow us to constrain the magnetic field strength to $\lesssim 1.5 \ \mu G$ under the leptonic scenario.
  \item The analysis of the \fermilat{} data above $100 \ \mathrm{MeV}$ revealed a point-like source 4FGL J2108.0+5155 towards the LHAASO J2108+5157 region. Its properties align with those reported in the 4FGL-DR3 catalog (\fermi{}). The measured spectrum shows a pulsar-like spectrum with a steep cutoff above 1 GeV. 
  \item We have reported a non-detection of LHAASO J2108+5157 using VERITAS data above 200 GeV. The spectral ULs of this observation are consistent with the WCDA spectrum \citep{Lhaaso2023catalogue} and constrain the emission model parameters in the 1-10 TeV range.
  \item With 2400 days of HAWC data, we have detected the source at a significance level of $7.5~\sigma$ above 300~GeV. Furthermore, an extended source with an extension of $0.21^{\circ} \pm 0.04^{\circ}$ is slightly favored. The best-fit position of the emission centroid in HAWC is at $0.06^{\circ} \pm 0.07^{\circ}$ offset from the position measured by KM2A.
  \item We provided a benchmark model for the multi-wavelength SED under the assumption that LHAASO J2108+5157 is a PWN powered by a (yet to be identified) gamma-ray pulsar \fermi{}. Under this leptonic scenario, the GeV detected by \fermilat{} is attributed to the pulsar emission, and the X-ray and TeV emission is explained by synchrotron radiation and inverse Compton scattering, respectively, of relativistic electrons accelerated within the PWN. The electron spectrum is described by a power law with a super-exponential cutoff ($\beta=2$), electron index $\alpha_e=1.6$, and cutoff energy $E_{e,cut}=196$ TeV.
\end{itemize}

This research is supported by grants from the U.S. Department of Energy Office of Science, the U.S. National Science Foundation and the Smithsonian Institution, by NSERC in Canada, and by the Helmholtz Association in Germany. This research used resources provided by the Open Science Grid, which is supported by the National Science Foundation and the U.S. Department of Energy's Office of Science, and resources of the National Energy Research Scientific Computing Center (NERSC), a U.S. Department of Energy Office of Science User Facility operated under Contract No. DE-AC02-05CH11231.
We acknowledge the excellent work of the technical support staff at the Fred Lawrence Whipple Observatory and at the collaborating institutions in the construction and operation of the instrument.

We acknowledge the support from: the US National Science Foundation (NSF); the US Department of Energy Office of High-Energy Physics; the Laboratory Directed Research and Development (LDRD) program of Los Alamos National Laboratory; Consejo Nacional de Ciencia y Tecnolog\'{i}a (CONACyT), M\'{e}xico, grants LNC-2023-117, 271051, 232656, 260378, 179588, 254964, 258865, 243290, 132197, A1-S-46288, A1-S-22784, CF-2023-I-645, c\'{a}tedras 873, 1563, 341, 323, Red HAWC, M\'{e}xico; DGAPA-UNAM grants IG101323, IN111716-3, IN111419, IA102019, IN106521, IN114924, IN110521 , IN102223; VIEP-BUAP; PIFI 2012, 2013, PROFOCIE 2014, 2015; the University of Wisconsin Alumni Research Foundation; the Institute of Geophysics, Planetary Physics, and Signatures at Los Alamos National Laboratory; Polish Science Centre grant, 2024/53/B/ST9/02671; Coordinaci\'{o}n de la Investigaci\'{o}n Cient\'{i}fica de la Universidad Michoacana; Royal Society - Newton Advanced Fellowship 180385; Gobierno de España and European Union - NextGenerationEU, grant CNS2023- 144099; The Program Management Unit for Human Resources \& Institutional Development, Research and Innovation, NXPO (grant number B16F630069); Coordinaci\'{o}n General Acad\'{e}mica e Innovaci\'{o}n (CGAI-UdeG), PRODEP-SEP UDG-CA-499; Institute of Cosmic Ray Research (ICRR), University of Tokyo. H.F. acknowledges support by NASA under award number 80GSFC21M0002. C.R. acknowledges support from National Research Foundation of Korea (RS-2023-00280210). We also acknowledge the significant contributions over many years of Stefan Westerhoff, Gaurang Yodh and Arnulfo Zepeda Dom\'inguez, all deceased members of the HAWC collaboration. Thanks to Scott Delay, Luciano D\'{i}az and Eduardo Murrieta for technical support.

The \xmm{} observation and data analysis are supported by NASA grant XMMNC22.

\bibliography{bibliography}{}
\bibliographystyle{aasjournal}



\end{document}